# Magnetocaloric Effect in the Symmetric Spin-1/2 Diamond Chain with Different Landé g - factors of the Ising and Heisenberg Spins

**Lucia Gálisová***

Department of Applied Mathematics and Informatics, Faculty of Mechanical Engineering, Technical University of Košice, Letná 9, 042 00 Košice, Slovak Republic

**Abstract:** The symmetric spin-1/2 Ising-Heisenberg diamond chain with different Landé g-factors of the Ising and Heisenberg spins is exactly solved by combining the generalized decoration-iteration transformation and transfer-matrix method. The ground state of the system and the magnetocaloric effect during the adiabatic (de)magnetization are particularly examined. It is evidenced that the considered mixed-spin diamond chain exhibits the enhanced magnetocaloric effect during the adiabatic (de)magnetization in the vicinity of field-induced phase transitions as well as in the zero-field limit when the frustrated phase constitutes the zero-field ground state. The cooling efficiency of the system depends on whether it is macroscopically degenerate in these parameter regions or not.

## 1. Introduction

Exactly solvable quantum spin chains belong to attractive issues of the statistical physics, because they offer a valuable insight into many unconventional physical phenomena [1]. One group of such systems are hybrid Ising-Heisenberg diamond chains composed of "classical" Ising and quantum Heisenberg spins, which provide the exact evidence of the existence of several novel and unexpected quantum states [2-4], intermediate magnetization plateaus in low-temperature magnetization curves [4-7], multi-peak structure in temperature dependencies of the specific heat [7,8], enhanced magnetocaloric effect during the adiabatic (de)magnetization [4,5,8], thermal entanglement [9], ect. However, immense theoretical interest focused on the mixed-spin Ising-Heisenberg diamond chains is not purpoless. Despite some simplifications, the mixed-spin Ising-Heisenberg model with diamond-chain geometry qualitatively reproduces the most important features of the real natural mineral single crystal $Cu_3(CO_3)_2(OH)_2$, known as azurite, such as the magnetization plateau at one-third of the saturation magnetization as well as the double-peak structure in temperature dependencies of the specific heat and zero-field susceptibility [10,11].

In this work, we will investigate the symmetric spin-1/2 Ising-Heisenberg diamond chain with different Landé g-factors of the Ising and Heisenberg spins. Beside the ground-state analysis, the main goal of this paper is to study the magnetocaloric effect during the adiabatic (de)magnetization of the considered mixed-spin model.

*** Corresponding author:** RNDr. Lucia Gálisová, PhD., ***Tel.:*** *+421-55-602-2228*
***E-mail address:*** *lucia.galisova@tuke.sk*

The paper is organized as follows. In Section 2, the most important steps of an exact analytical treatment for the model under investigation will be presented. Section 3 deals with a discussion of the most interesting numerical results for the ground state and the adiabatic (de)magnetization of the system. Finally, some concluding remarks will be drawn in Section 4.

## 2. Model and its Exact Solution

Let us consider an one-dimensional lattice composed of $N$ inter-connected diamonds as is illustrated in Fig. 1. In this figure, black circles denote nodal lattice sites occupied by the Ising spins $\sigma = 1/2$ and red circles label decorating lattice sites occupied by the Heisenberg spins $S = 1/2$. Taking into account a "classical" nature of the Ising spins, which represent a barrier for local quantum fluctuations induced by the Heisenberg spins, the total Hamiltonian of the considered spin-1/2 Ising-Heisenberg diamond chain may be written as a sum of $N$ commuting block Hamiltonians $\hat{H}_k$:

$$\hat{H} = \sum_{k=1}^{N} \hat{H}_k, \tag{1}$$

where each block Hamiltonian $\hat{H}_k$ involves all the interaction terms that belong to the $k$th primitive cell:

$$\hat{H}_k = J_H \left[ \Delta \left( \hat{S}^x_{3k-1}\hat{S}^x_{3k} + \hat{S}^y_{3k-1}\hat{S}^y_{3k} \right) + \hat{S}^z_{3k-1}\hat{S}^z_{3k} \right] + J_I \left( \hat{S}^z_{3k-1} + \hat{S}^z_{3k} \right)\left( \sigma^z_{3k-2} + \sigma^z_{3k+1} \right) - \mu_B g^z_H B \left( \hat{S}^z_{3k-1} + \hat{S}^z_{3k} \right) - \frac{\mu_B g^z_I B}{2}\left( \sigma^z_{3k-2} + \sigma^z_{3k+1} \right). \tag{2}$$

In above, $\hat{S}^\alpha_k$ ($\alpha = x, y, z$) and $\sigma^z_k$ denote spatial components of the spin-1/2 operators, the parameter $J_H$ labels the anisotropic XXZ interaction between the nearest-neighbouring Heisenberg spins, $\Delta$ is an exchange anisoptropy in this interaction and the parameter $J_I$ denotes the Ising interaction between the Heisenberg spins and their nearest Ising neighbours. Finally, the last two terms determine the Zeeman's energy of the Heisenberg and Ising spins placed in the magnetic field $B$ oriented along the $z$th axis, respectively. The quantity $\mu_B$ is a Bohr magneton and $g^z_H, g^z_I$ label $z$th components of Landé g-factors corresponding to the Heisenberg and the Ising spins, respectively.

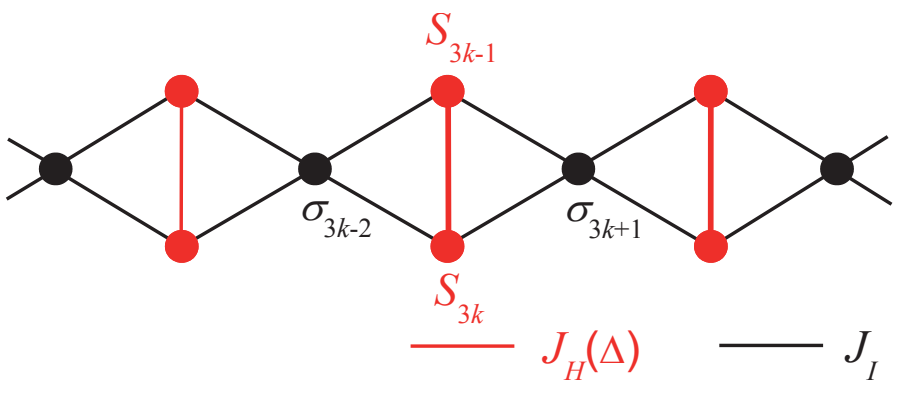

Fig. 1: *A part of the hybrid spin-1/2 Ising-Heisenberg diamond chain. The black (red) circles label lattice positions of the Ising (Heisenberg) spins.*

### 2.1 Partition Function

The most important part of our calculations is an evaluation of the partition function for the investigated model. The commutation relation between different block Hamiltonians $[\hat{H}_k, \hat{H}_l] = 0$ $(k \neq l)$ allows to partially factorize the partition function of the system into a product of the block partition functions $Z_k$:

$$Z = \sum_{\{\sigma_i\}} \prod_{k=1}^{N} \mathrm{Tr}_k \exp\left(-\beta \hat{H}_k\right) = \sum_{\{\sigma_i\}} \prod_{k=1}^{N} Z_k, \tag{3}$$

where $\beta = 1/T$, $T$ is the absolute temperature (we set $k_B$=1), the symbol $\Sigma_{\{\sigma_i\}}$ denotes a summation over all available configurations of the Ising spins $\{\sigma_i\}$ and the symbol $\mathrm{Tr}_k$ labels a trace over spin degrees of freedom of the Heisenberg spin pair from $k$th diamond cell. The block partition function $Z_k$ can be subsequently acquired by a diagonalization of the block Hamiltonian (2). For this purpose, it is useful to pass to the matrix representation of the block Hamiltonian (2) in the orthogonal basis constructed by the eigenstates of the operator $\hat{S}^z_{3k-1}\hat{S}^z_{3k}: \left\{ |\uparrow,\uparrow\rangle_{3k-1,3k}, |\uparrow,\downarrow\rangle_{3k-1,3k}, |\downarrow,\uparrow\rangle_{3k-1,3k}, |\downarrow,\downarrow\rangle_{3k-1,3k} \right\}$, where the relevant ket vectors $|\uparrow(\downarrow),\uparrow(\downarrow)\rangle_{3k-1,3k} = |\uparrow(\downarrow)\rangle_{3k-1} |\uparrow(\downarrow)\rangle_{3k}$ correspond to the appropriate combinations of the spin states $\hat{S}^z_{3k-1} = \pm 1/2$ and $\hat{S}^z_{3k} = \pm 1/2$ of two Heisenberg spins at (3$k$-1)th and 3$k$th lattice sites, respectively. The diagonalization of the block Hamiltonian (2) transcribed in the matrix representation yields the four different energy eigenvalues:

$$E_{1,2}\left(\sigma^z_{3k-2}, \sigma^z_{3k+1}\right) = \frac{J_H}{4} \pm J_I \left( \sigma^z_{3k-2} + \sigma^z_{3k+1} \right) \mp \mu_B g^z_H B - \frac{\mu_B g^z_I B}{2}\left( \sigma^z_{3k-2} + \sigma^z_{3k+1} \right),$$

$$E_{3,4}\left(\sigma^z_{3k-2}, \sigma^z_{3k+1}\right) = \frac{\pm J_H \Delta}{2} - \frac{J_H}{4} - \frac{\mu_B g^z_I B}{2}\left( \sigma^z_{3k-2} + \sigma^z_{3k+1} \right). \tag{4}$$

The set of eigenvalues (4) can directly be used to obtain the expression for the block partition function $Z_k$, which implies the possibility of performing the generalized decoration-iteration mapping transformation [12,13]:

$$Z_k = \sum_{j=1}^{4} \exp\left(-\beta E_j\right) = 2\exp\left[\frac{\beta\mu_B g_I^z B}{2}\left(\sigma_{3k-2}^z + \sigma_{3k+1}^z\right)\right]$$
$$\times\left\{\exp\left(\frac{-\beta J_H}{4}\right)\cosh\left[\beta J_I\left(\sigma_{3k-2}^z + \sigma_{3k+1}^z\right) \mp \beta\mu_B g_H^z B\right]\right.$$
$$\left. + \exp\left(\frac{\beta J_H}{4}\right)\cosh\left(\frac{\beta J_H \Delta}{2}\right)\right\}$$
$$= A\exp\left[\beta J_{eff}\sigma_{3k-2}^z \sigma_{3k+1}^z + \frac{\beta B_{eff}}{2}\left(\sigma_{3k-2}^z + \sigma_{3k+1}^z\right)\right]. \tag{5}$$

From the physical point of view, the mapping transformation (5) effectively removes all the interaction terms associated with a couple of Heisenberg spins from *k*th diamond cell and replaces them by the effective interaction $J_{eff}$ and the effective field $B_{eff}$ acting on the remaining Ising spins $\sigma_{3k-2}$ and $\sigma_{3k+1}$. Of course, the transformation relation (5) must hold for all possible spin combinations of the Ising spins $\sigma_{3k-2}$ and $\sigma_{3k+1}$. This "self-consistency" condition unambiguously determines the unknown mapping parameters $A$, $J_{eff}$ and $B_{eff}$:

$$A = 2\sqrt[4]{V_+ V_- V_0^2},$$
$$J_{eff} = T\left(\ln V_+ + \ln V_-\right) - 2T\ln V_0,$$
$$B_{eff} = \mu_B g_I^z B - T\left(\ln V_+ - \ln V_-\right), \tag{6}$$

where $V_\pm$ and $V_0$ are independent expressions to be obtained by substituting four possible configurations of the Ising spins $\sigma_{3k-2}$, $\sigma_{3k+1}$ into the block partition function (5):

$$V_\pm = \exp\left(\frac{-\beta J_H}{4}\right)\cosh\left(\beta J_I \pm \beta\mu_B g_H^z B\right)$$
$$+ \exp\left(\frac{\beta J_H}{4}\right)\cosh\left(\frac{\beta J_H \Delta}{2}\right),$$
$$V_0 = \exp\left(\frac{-\beta J_H}{4}\right)\cosh\left(\beta\mu_B g_H^z B\right)$$
$$+ \exp\left(\frac{\beta J_H}{4}\right)\cosh\left(\frac{\beta J_H \Delta}{2}\right). \tag{7}$$

The substitution of the transformation relation (5) into the Eq. (3) yields the equality

$$Z = A^N Z_{Ising}, \tag{8}$$

which establishes a simple mapping relationship between the partition function $Z$ of the spin-1/2 Ising-Heisenberg diamond chain and the partition function $Z_{Ising}$ of the uniform spin-1/2 Ising linear chain with the effective interaction $J_{eff}$ between the nearest spin neighbours and the effective magnetic field $B_{eff}$. In fact, Eq. (8) completes the exact calculation of the partition function of the investigated mixed-spin model, since the partition function of the uniform spin-1/2 Ising chain can be evaluated within the framework of the transfer-matrix method [14]:

$$Z_{Ising} = \exp\left(\frac{N\beta J_{eff}}{4}\right)\times$$
$$\left[\cosh\left(\frac{\beta B_{eff}}{2}\right) + \sqrt{\sinh^2\left(\frac{\beta B_{eff}}{2}\right) + \exp\left(-\beta J_{eff}\right)}\right]^N. \tag{9}$$

### 2.2 Helmholtz Free Energy and Entropy

Exact results for all thermodynamic quantities of the considered mixed-spin diamond chain follow straightforwardly from the mapping relation (8). Actually, the Helmholtz free energy of the model may be evaluated from the formula:

$$F = -T\ln Z = -T\ln Z_{Ising} - NT\ln A, \tag{10}$$

which subsequently allows a direct calculation, for instance, the entropy $S$ of the model:

$$S = -\left(\frac{\partial F}{\partial T}\right)_B. \tag{11}$$

## 3. Results and Discussion

In this section, let us proceed to an analysis of the ground state and adiabatic demagnetization process of the symmetric spin-1/2 Ising-Heisenberg diamond chain with different *z*th components of Landé g-factors $g_I^z$ and $g_H^z$ for the Ising and Heisenberg spins, respectively. Even though all results derived in the preceding section hold irrespective of the nature of exchange interactions between the nearest-neighbouring spins, we will assume both the exchange parameters $J_I$ and $J_H$ to be antiferromagnetic ($J_I > 0$, $J_H > 0$). One can expect that the magnetic behaviour of the model with the antiferromagnetic exchange interactions placed in the longitudinal magnetic field should be much more interesting compared with its ferromagnetic counterpart. To reduce the numer of interaction parameters in the system, we will further assume the fixed exchange anisotropy parameter $\Delta=1$ and the fixed *z*th component of Landé g-factor of Heiseberg spins $g_H^z = 2$ in contrast to the variable *z*th component of Landé g-factor of Ising spins $g_I^z > g_H^z$.

### 3.1 Ground State

First, let us begin with the discussion of the possible spin configuration of the investigated diamond chain at zero temperature. Typical ground-state phase diagrams in the $J_H / J_I$ - $\mu_B B / J_I$ plane including all possible ground states are depicted in Fig. 2 for two different sets of Landé g-factors of Ising spins: $g_I^z \leq 4$ [Fig. 2(a)] and $g_I^z > 4$ [Fig. 2(b)]. As one can see from Fig. 2(a), three different phases $\mathrm{FRI}_1$, SPP and FRU appear in the ground state due to a mutual interplay between the exchange interactions $J_I$, $J_H$ and the applied magnetic field $B$ when assuming $g_I^z \leq 4$. The respective ground states can be unambiguously characterized by the following eigenvectors and eigenenergies:

$$|\mathrm{FRI}_1\rangle = \prod_{k=1}^{N} |\downarrow\rangle_{3k-2} \otimes |\uparrow,\uparrow\rangle_{3k\text{-}1,3k},$$
$$E_{\mathrm{FRI}_1} = \frac{N}{4}\left(J_H - 4J_I - 8\mu_B B + 2\mu_B g_I^z B\right); \quad (12)$$

$$|\mathrm{SPP}\rangle = \prod_{k=1}^{N} |\uparrow\rangle_{3k-2} \otimes |\uparrow,\uparrow\rangle_{3k\text{-}1,3k},$$
$$E_{\mathrm{SPP}} = \frac{N}{4}\left(J_H + 4J_I - 8\mu_B B - 2\mu_B g_I^z B\right); \quad (13)$$

$$|\mathrm{FRU}\rangle = \begin{cases} \prod_{k=1}^{N} |\downarrow(\uparrow)\rangle_{3k-2} \otimes |\varphi\rangle_k, & B = 0, \\ \prod_{k=1}^{N} |\uparrow\rangle_{3k-2} \otimes |\varphi\rangle_k, & B \neq 0, \end{cases}$$
$$|\varphi\rangle_k = \frac{1}{\sqrt{2}}\left(|\uparrow,\downarrow\rangle_{3k\text{-}1,3k} - |\downarrow,\uparrow\rangle_{3k\text{-}1,3k}\right),$$
$$E_{\mathrm{FRU}} = -\frac{N}{4}\left(J_H + 2J_H + 2\mu_B g_I^z B\right). \quad (14)$$

In above, the ket vectors $|\uparrow(\downarrow)\rangle$ after relevant products describe the spin states $\sigma^z = \pm 1/2$ and $S^z = \pm 1/2$ of the Ising and Heisenberg spins, respectively. As one clearly sees from Eqs. (12) and (13), the phase $\mathrm{FRI}_1$ represents the semi-classical ferrimagnetic state with opposite orientation of the nodal Ising spins with respect to the fully polarized Heisenberg dimers and the phase SPP is the saturated paramagnetic state characterized by a full alignment of the Ising and Heisenberg spins into the field direction. Evidently, both the phases $\mathrm{FRI}_1$ and SPP exhibit spin arrangements commonly observed also in the pure Ising systems. By contrast, the FRU ground state has a character of quantum frustrated phase. Actually, when the external magnetic field $B$ is zero, then the nodal Ising spins are completely free to flip in this phase owing to the spin frustration, which arises out of quantum entanglement of the decorating Hesenberg spins described by the antisymmetric wave function $\left(|\uparrow,\downarrow\rangle - |\downarrow,\uparrow\rangle\right)/\sqrt{2}$ [see Eq. (14)]. On the other hand, all Ising spins are fully polarized into the field direction in the FRU phase, whenever the external magnetic field $B$ is non-zero. All three phases $\mathrm{FRI}_1$, SPP, FRU meet at the triple point $\mathrm{T}_1$ with the coordinates

$$\mathrm{T}_1\left[\frac{J_H}{J_I}, \frac{\mu_B B}{J_I}\right] = \left[\frac{4 - g_I^z}{g_I^z}, \frac{2}{g_I^z}\right], \quad (15)$$

which gradually moves towards lower values of the iteraction ratio $J_H/J_I$ and the magnetic field $\mu_B B/J_I$ with the increasing g-factor $g_I^z$ of the Ising spins until a direct field-induced transition between the $\mathrm{FRI}_1$ and SPP phases observable along the line $\mu_B B / J_I = 2 / g_I^z$ completely vanishes on account of presence of the FRU phase.

More simple situation emerges when assuming $g_I^z > 4$ [see Fig. 2(b)]. In this particular case, the ground-state phase diagram consists of two previously reported phases SPP and FRU described by Eqs. (13) and (14), respectively, and one new semi-classical ferrimagnetic phase $\mathrm{FRI}_2$ characterized by the following eigenvector and eigenenergy:

$$|\mathrm{FRI}_2\rangle = \prod_{k=1}^{N} |\uparrow\rangle_{3k-2} \otimes |\downarrow,\downarrow\rangle_{3k\text{-}1,3k},$$
$$E_{\mathrm{FRI}_2} = \frac{N}{4}\left(J_H - 4J_I + 8\mu_B B - 2\mu_B g_I^z B\right). \quad (16)$$

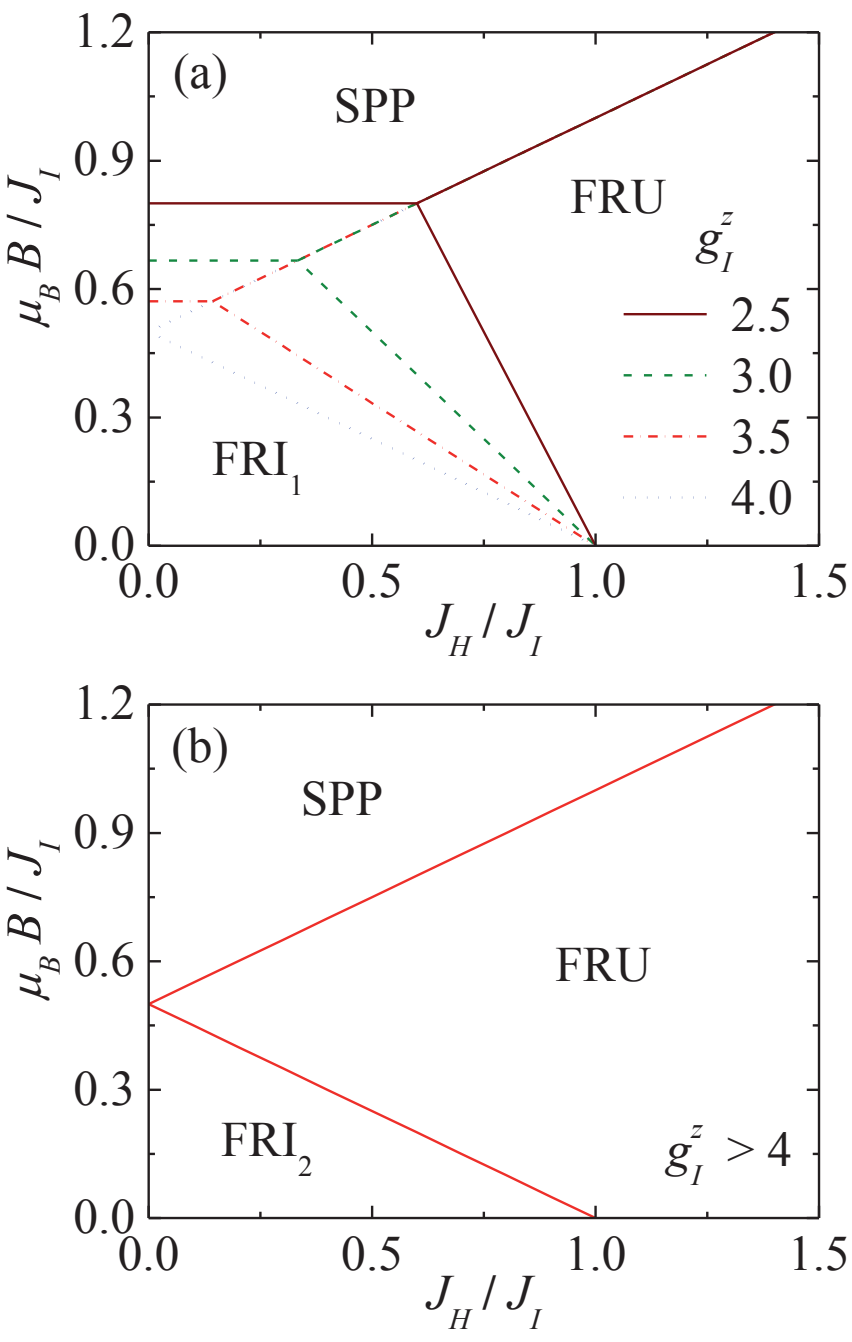

**Fig. 2:** *The ground-state phase diagram in the $J_H / J_I$ - $\mu_B B/J_I$ plane for the spin-1/2 Ising-Heisenberg diamond chain with two different sets of Landé g-factors of Ising spins: (a) $g_I^z \leq 4$ and (b) $g_I^z > 4$.*

It is clear from Eq. (16) that the spin arrangement of FRI$_2$ phase is precisely opposite compared to the one of the FRI$_1$ phase: the Ising spins are fully polarized towards the field direction, while the Heisenberg spins are aligned antiparallel with respect to their Ising neigbours. All three phases FRI$_2$, FRU and SPP meet at the triple point T$_2$ with constant coordinates

$$\mathrm{T}_2\left[\frac{J_H}{J_I}, \frac{\mu_B B}{J_I}\right] = \left[0, \frac{1}{2}\right]. \quad (17)$$

The field-induced boundaries separating relevant phases remain unchanged for any value of g-factor $g_I^z > 4$.

### 3.2 Enhanced Magnetocaloric Effect

Now, let us examine magnetocaloric properties of the considered spin-1/2 Ising-Heisenberg diamond chain in its classical interpretation as an adiabatic change of the temperature under the external magnetic field variation. Fig. 3 illustrates a density plot of the entropy in the $J_H /J_I$ -$\mu_B B/J_I$ plane at the constant temperature $T / J_I = 0.05$ by assuming the fixed g-factor of Ising spins $g_I^z = 2.5$. As one can see from this figure, the residual entropy can be observed along the field-induced phase transitions between the ground states FRI$_1$, SPP and FRU, SPP, as well as in a zero-field limit for any interaction ratios $J_H /J_I \geq 1.0$. This finding indicates a macroscopic degeneracy of the system in these parameter regions. By contrast, the field-induced transition between the FRI$_1$ and SPP phases is characterized by zero contribution of the entropy in the thermodynamic limit $N \to \infty$, which implies that the system under investigation is not macroscopically degenerate at critical fields corresponding to this particular phase transition.

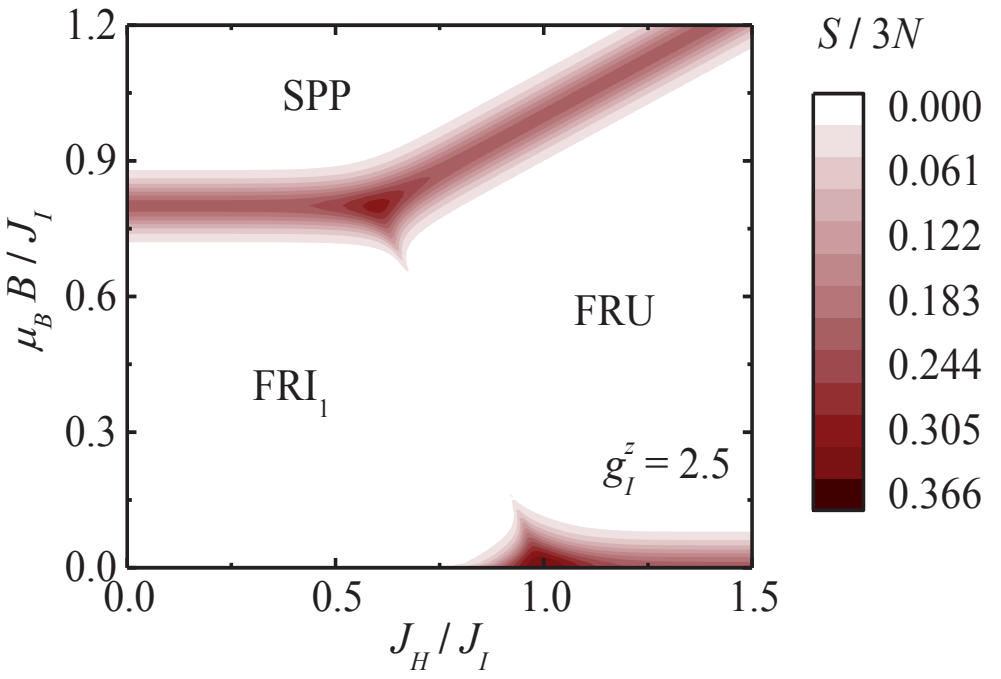

**Fig. 3:** *A density plot of the entropy in the $J_H / J_I$ - $\mu_B B / J_I$ plane at the constant temperature $T / J_I = 0.05$ and $g_I^z = 2.5$.*

The non-zero residual entropy at critical fields corresponding to the field-induced phase transitions FRI$_1$-SPP, FRU-SPP and in the zero-field limit for $J_H /J_I \geq 1.0$ gives rise to an enhanced magnetocaloric effect accompanied by a relatively fast and efficient cooling of the system to zero temperature during the adiabatic (de)magnetization, as is demonstrated in Fig. 4 which shows a density plot of the entropy as a function of the magnetic field and the temperature together with isentropic changes of the temperature upon varying the magnetic field. Note that parameter $g_I^z$ takes the same values as in Fig. 3 and the values of the interaction ratio $J_H / J_I$ are chosen so as to cover all possible ground-state phase transitions as well as the case when the zero-field ground state is macroscopically degenerate. The adiabatic (de)magnetization related to the field-induced phase transitions between the phases FRI$_1$ and SPP can be analyzed from results shown in Fig. 4(a). Evidently, the model under investigation is rapidly cooled down to zero temperature in a vicinity of the phase transition FRI1-SPP whenever the entropy is set close enough to the value $S/3N = \ln 2^{1/3} \approx 0.231$, but the temperature tends towards a finite value as the magnetic field vanishes. On the other hand, the enhanced magnetocaloric effect during the adiabatic (de)magnetization can be observed both at the zero field and around the critical field corresponding to the field-induced phase transition FRU-SPP if a mutual competition between the exchange interactions $J_H$ and $J_I$ drives the system into the disordered FRU ground state in the zero-field limit [see Fig. 4(c)]. Under this circumstance, the most abrupt isentropic temperature drop to the absolute zero is achieved when the entropy is again set sufficiently close to the value $S/3N = \ln 2^{1/3} \approx 0.231$. Different behaviour can be observed in the vicinity of the critical field corresponding to the phase transition $FRI_1$-FRU. Provided that the entropy normalized per spin is kept very close to the zero value, the temperature of the system sharply falls, but it reaches just some non-zero (even if relatively small) value when the magnetic field approaches the critical value corresponding to the phase transition between the FRI1 and FRU ground states [see the curves plotted for $S/3N = 0.001$ and $0.01$ in Fig. 4(b)]. As expected, the observed local minimum in the magnetic field dependence of the temperature disappears upon an increase of the entropy during the adiabatic process.

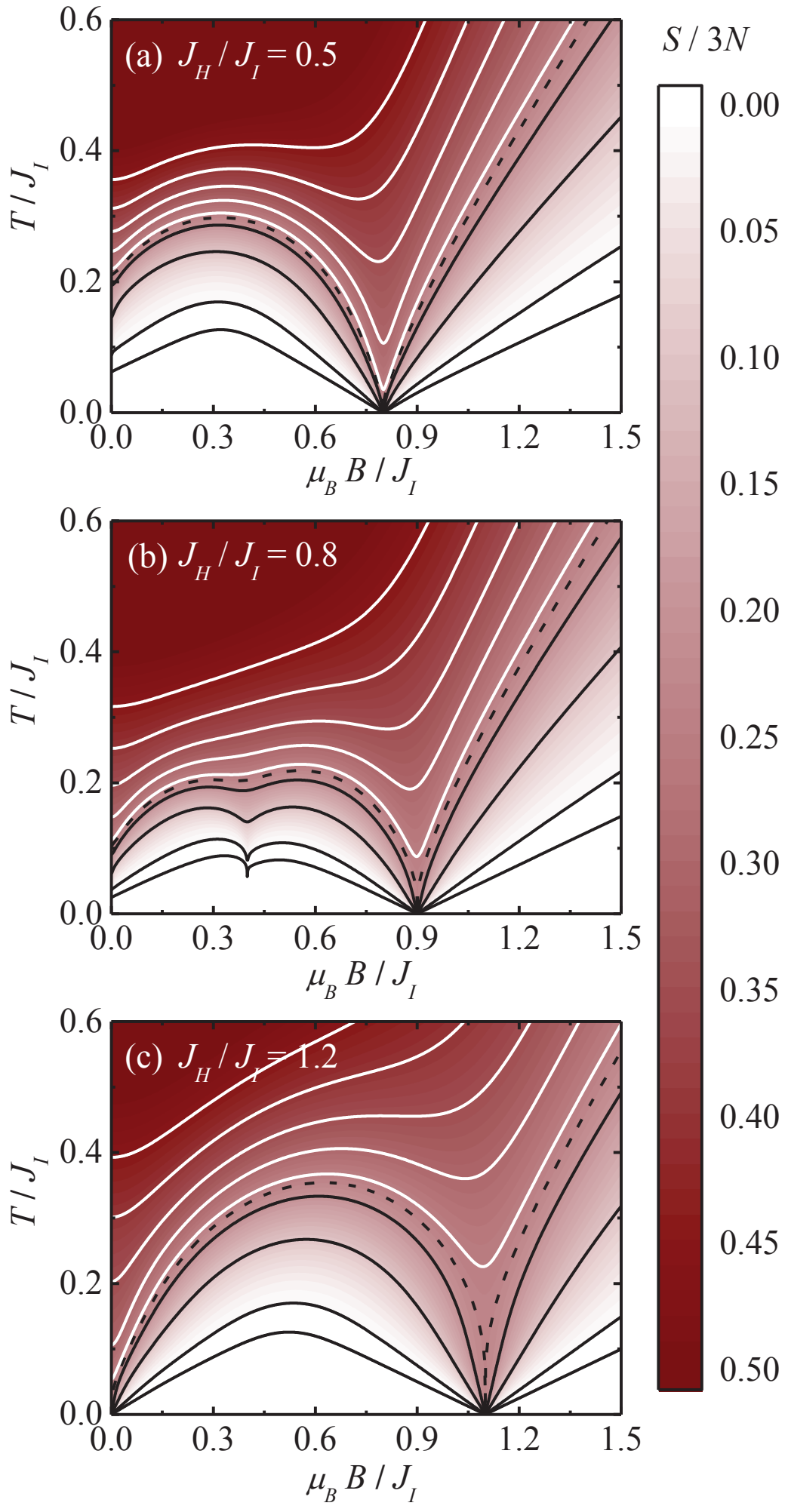

**Fig. 4:** *A density plot of the entropy as a function of the magnetic field and temperature for* $g_I^z = 2.5$ *and (a)* $J_H / J_I = 0.05$*, (b)* $J_H / J_I = 0.8$*, (c)* $J_H / J_I = 1.2$*. The displayed curves correspond to isentropic lines:* $S/3N = 0.001, 0.01, 0.1, 0.2$ *(black solid curves),* $S/3N = \ln 2^{1/3}$ *(black dashed curve),* $S/3N = 0.25, 0.3, 0.35, 0.4, 0.45$ *(white solid curves).*

For complete analysis of the magnetocaloric effect in the symmetric spin-1/2 Ising-Heisenberg diamond chain with different Landé g-factors, let us also investigate magnetocaloric properties of the system with g-factors of the Ising spins $g_I^z > 4$. Our findings are displayed in Figs. 5 and 6. Fig. 5 illustrates a density plot of the entropy in the $J_H/J_I$ - $\mu_B B/J_I$ plane at the constant temperature $T/J_I = 0.05$ and $g_I^z > 4$. It is obvious from this figure that the system is macroscopically degenerate at critical fields related to the field-induced phase transitions $FRI_2$-SPP and FRU-SPP, as well at the zero field, where the FRU phase constitutes the ground state. As a result, the non-zero residual entropy $S/3N = \ln 2^{1/3} \approx 0.231$ is accumulated at zero temperature in these regions.

To investigate the magnetocaloric effect during the adiabatic (de)magnetization nearby the aforementioned parameter regions in more detail, the density plot of the entropy as a function of the magnetic field and temperature is depicted in Fig. 6 for $g_I^z = 5$ and two different values of interaction ratio $J_H / J_I = 0.5$ and $1.2$. Isentropic changes of the temperature upon varying the magnetic field can be identified in Fig. 6 as solid and dashed curves. Obviously, the most fast and rather efficient cooling of the system during the adiabatic process can be achieved only if the entropy sufficienty close to the value $S/3N = \ln 2^{1/3} \approx 0.231$.

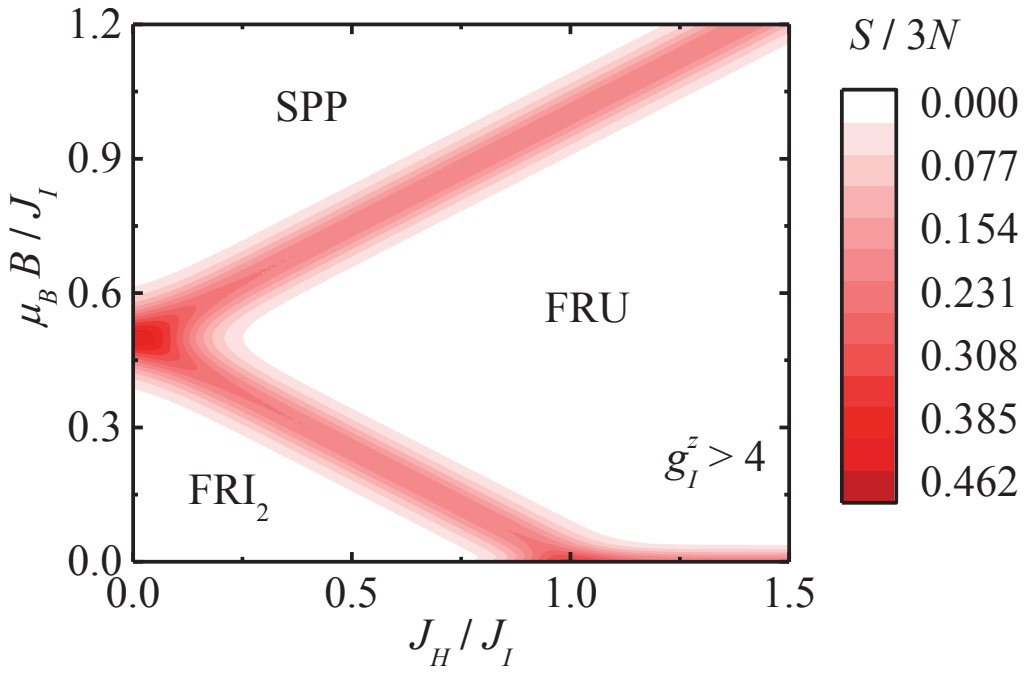

Fig. 5: *A density plot of the entropy in the* $J_H / J_I$ - $\mu_B B / J_I$ *plane at the constant temperature* $T / J_I = 0.05$ *and* $g_I^z > 4$.

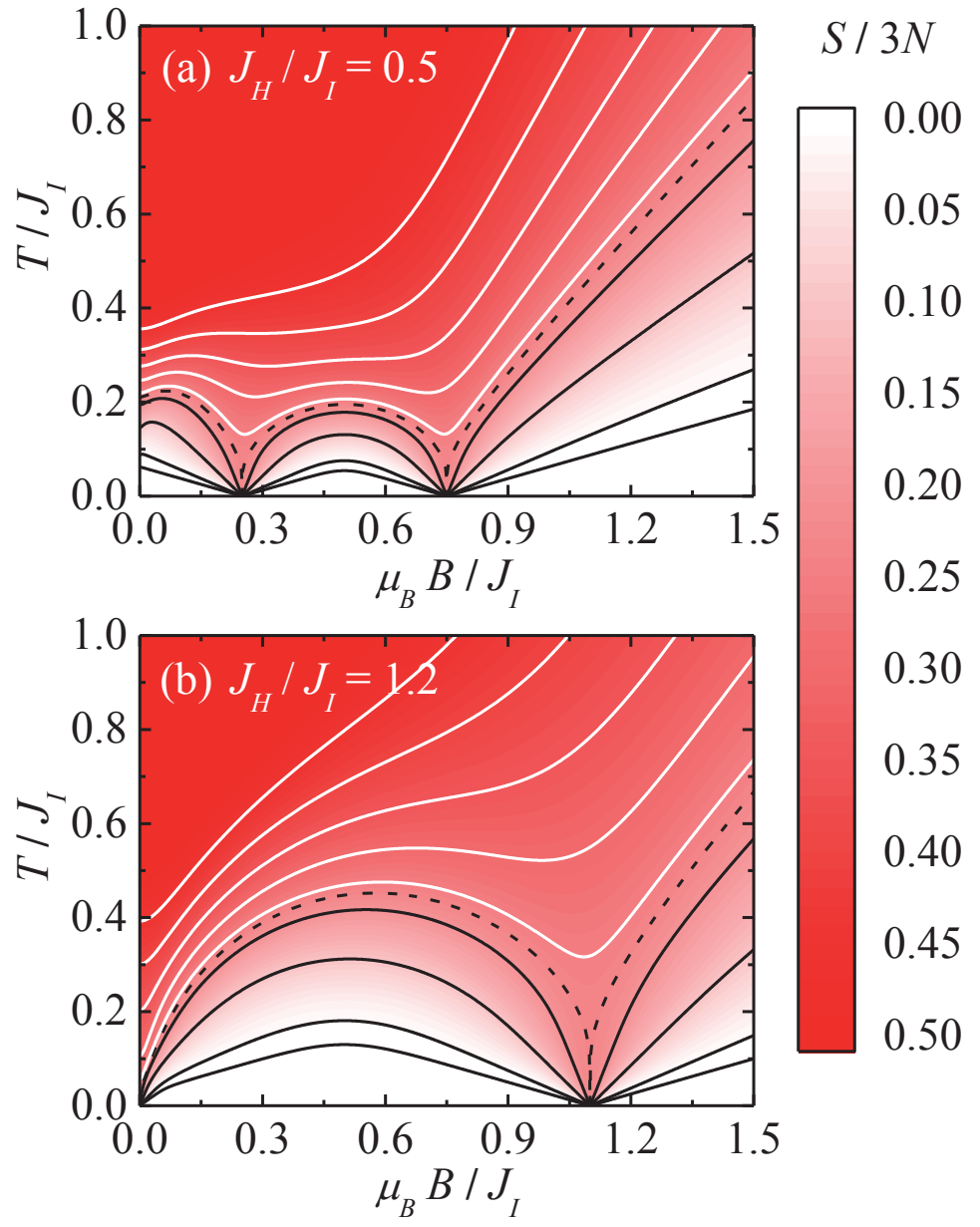

Fig. 6: *A density plot of the entropy as a function of the magnetic field and temperature for* $g_I^z = 5$ *and (a)* $J_H / J_I = 0.05$, *(b)* $J_H / J_I = 1.2$. *The displayed curves correspond to isentropic lines:* $S/3N = 0.001, 0.01, 0.1, 0.2$ *(black solid curves),* $S/3N = \ln 2^{1/3}$ *(black dashed curve),* $S/3N = 0.25, 0.3, 0.35, 0.4, 0.45$ *(white solid curves).*

Under this circumstance, the temperature of the system falls rather quickly to the zero value, if the external magnetic field is approaching either the critical value related to phase transitions between the ground states $FRI_2$, SPP and FRU, SPP, or it tends towards zero.

## 4. Conclusions

In the present paper, we have studied the ground state and magnetocaloric effect of the symmetric spin-1/2 Ising–Heisenberg diamond chain with different Landé g-factors of the Ising and Heisenberg spins, which is exactly solvable by combining the generalized decoration-iteration transformation and transfer-matrix method. We have evidenced that the ground-state phase diagram of the investigated diamond chain is formed by quantum frustrated phase, classical saturated phase and either semi-classical ferrimagnetic phase with the opposite orientation of nodal Ising spins with respect to the fully polarized Heisenberg dimers, or semi-classical ferrimagnetic phase, where the nodal Ising spins are fully polarized into the field direction and the Heisenberg spins are aligned antiparallel with respect to their Ising neighbours depending on whether the g-factor of the Ising spins is $g_I^z \leq 4$ or $g_I^z > 4$, respectively. We have also illustrated the density plots of the entropy in the $J_H/J_I$-$\mu_B B/J_I$ plane at the constant temperature and the adiabatic temperature changes upon varying the external magnetic field for both particular cases of $g_I^z$. It has been demonstrated that the system exhibits an enhanced magnetocaloric effect during the adiabatic (de)magnetization in the vicinity of field-induced phase transitions as well as in the zero-field limit if the frustrated phase constitutes the zero-field ground state. However, the cooling efficiency during the adiabatic (de)magnetization basically depends on whether the system is macroscopically degenerate in these parameter regions or not.

Finally, it should be mention that the considered symmetric Ising-Heisenberg diamond chain with different Landé g-factors of the Ising and Heisenberg spins, thanks to its simplicity, enables the rigorous analysis of the thermodynamic quantities, which may provide a deeper insight e.g. into the magnetization scenario as well as thermal dependencies of the specific heat and magnetic susceptibility of the model. In this direction will continue our next work.

## 5. Acknowledgments

*This work was financially supported by the grant of the Slovak Research and Development Agency under the contract No. APVV-0097-12.*

## Biographical notes

*__RNDr. Lucia Gálisová, PhD.__ (born in 1981) received PhD degree in General Physics and Mathematical Physics at the P. J. Šafárik University in Košice in 2008. At present, she works as a senior lecturer at the Department of Applied Mathematics and Informatics of the Faculty of Mechanical Engineering of Technical University of Košice. Her research interest is specialized in the theoretical investigation of magnetic properties of exactly solvable lattice-statistical models. She is co-author of 27 scientific publications registered in the Current Contents database, for which more than 80 citations were recorded in the citation databases SCI, WoS and SCOPUS.*